\documentclass[aps,prl,twocolumn,superscriptaddress]{revtex4}
\usepackage{amsmath}
\usepackage{amssymb}

\usepackage{amsmath,amssymb}
\usepackage[usenames]{color}
\usepackage{mathbbol}
\usepackage{amssymb}
\usepackage{grffile}
\usepackage[pdftex]{graphicx}
\usepackage{amsmath, amstext, amssymb, amsfonts, amsxtra}
\usepackage{textcomp}
\usepackage{xspace} 
\usepackage[colorlinks]{hyperref}

\DeclareSymbolFontAlphabet{\amsmathbb}{AMSb}

\newcommand{\aop}{\hat{a}}
\newcommand{\adop}{\hat{a}^{\dagger}}
\newcommand{\cop}{\hat{c}}
\newcommand{\cdop}{\hat{c}^{\dagger}}

\newcommand{\nop}{\hat{n}} 
\newcommand{\nbar}{\bar{n}}

\newcommand{\Hop}{\hat{H}}
\newcommand{\rhop}{\hat{\rho}}

\newcommand{\Dop}{\mathcal{D}}
\newcommand{\Pop}{\mathcal{P}}

\newcommand{\im}{{\rm i}}   
\newcommand{\hc}{{\rm H.c.}} 
\newcommand{\tr}{{\rm tr}}

\newcommand{\pare}[1]{\left(#1 \right)}
\newcommand{\spare}[1]{\left[#1 \right]}
\newcommand{\gpare}[1]{\left\{#1 \right\}}
 
\newcommand{\ga}{\gamma} 
\newcommand{\la}{\langle} 
\newcommand{\ra}{\rangle} 
   
\newcommand{\Cpm}{C_{+-}}

\newcommand{\bonn}{HISKP, University of Bonn, Nussallee 14-16, 53115 Bonn, Germany} 
\newcommand{\sutd}{Engineering Product Development Pillar, Singapore University of Technology and Design, 8 Somapah Road, 487372 Singapore}
\newcommand{\tng}{TNG Technology Consulting GmbH, Betastrasse 13a, 85774 Unterf\"ohring, Germany}

\begin{document}

\title{Light-cone and diffusive propagation of correlations in a many-body dissipative system}

\author{Jean-S\'ebastien Bernier}
\affiliation{\bonn} 
\author{Ryan Tan}
\affiliation{\sutd}
\author{Lars Bonnes}
\affiliation{\tng}
\author{Chu Guo}
\affiliation{\sutd} 
\author{Dario Poletti}
\affiliation{\sutd} 
\author{Corinna Kollath}
\affiliation{\bonn} 

\begin{abstract}
We analyze the propagation of correlations after a sudden interaction change in a strongly
interacting quantum system in contact with an environment. In particular, we consider an interaction quench
in the Bose-Hubbard model, deep within the Mott-insulating phase, under the effect of dephasing.
We observe that dissipation effectively speeds up the propagation of single-particle
correlations while reducing their coherence. In contrast, for two-point density correlations,
the initial ballistic propagation regime gives way to diffusion at intermediate times.
Numerical simulations, based on a time-dependent matrix product state algorithm,
are supplemented by a quantitatively accurate fermionic quasi-particle approach providing an intuitive
description of the initial dynamics in terms of holon and doublon excitations.
\end{abstract}

\date{\today}
\maketitle


In the last years, considerable experimental efforts have been devoted to dynamically
generate complex states and monitor their evolution. Ultrafast optical pulses
were used to photo-induce phase transitions in strongly interacting
materials~\cite{BasovDressel2011, Orenstein2012, ZhangAveritt2014, GiannettiMihailovic2016}
while similar successes were
reported for ultracold atoms using time-dependent
electromagnetic fields~\cite{BlochZwerger2008, PolkovnikovVengalattore2011}.
Despite these remarkable advances, the theoretical principles behind the
non-equilibrium dynamics of strongly correlated matter are still far from being understood.

One of the salient questions relates to the propagation of correlations in complex systems.
Providing an answer to this question would help understand how phase transitions can be dynamically induced.
For example, to tune a system from a disordered to an ordered phase, such as into a
superconducting or a N\'eel phase, one needs to establish how the correlations associated
to a given order parameter build up. Understanding how correlations propagate following a perturbation
would also provide major insights into the intricate mechanisms at play in complex systems. 

\begin{figure}
\includegraphics[width=0.9\columnwidth]{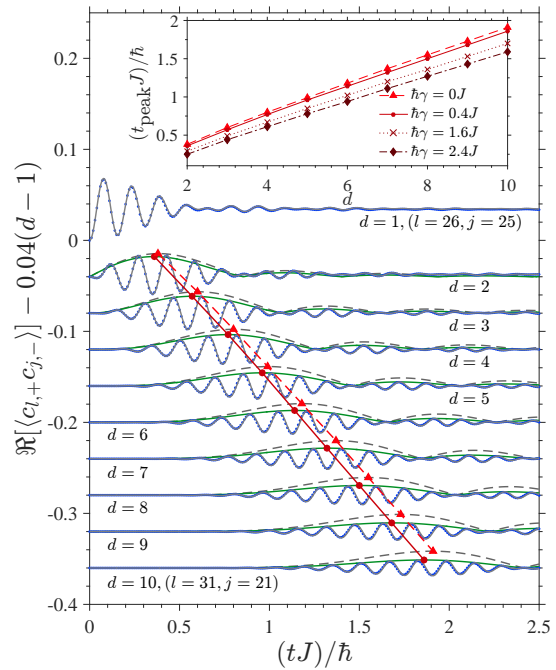}
\caption{(color online) Evolution of the real part of $\Cpm(d,t)$ for various distances following a
  sudden interaction change from an atomic Mott insulator to $U=40J$ in the presence of an environment with
  coupling $\hbar\gamma = 0.4J$ (the correlations are shifted by $-0.04(d-1)$ to improve readability).   
  Full blue circles: t-MPS results; full grey line: analytical
  solution; full green line: envelope of the analytical solution; dashed grey line: envelope of the analytical
  solution for $\gamma = 0$; red circles and triangles: position of the first maximum of $\Cpm$ for $\hbar\gamma = 0.4J$ and $0$. 
  Inset: time at which the first maximum of $\Cpm$ reaches a given distance for various dissipation strengths (obtained
  within the quasi-particle picture using Eq.~(\ref{eq:cpmsol})).
  \label{fig:Fig1} }
\end{figure}

Forays into the correlation dynamics of isolated systems have been made over the last decades. 
In a seminal work, Lieb and Robinson~\cite{LiebRobinson1972} demonstrated that for quantum spin systems
described by local Hamiltonians the propagation of correlations is bound by an effective light-cone.
This effect has for consequence that building up a given order parameter requires
some time as the correlations need to set in.  An elegant picture illustrating this
effect was developed in \cite{CalabreseCardy2006} using conformal field theory and exactly solvable models.
Within this picture, when a system is
quenched to an Hamiltonian close to a quantum critical point, the propagation of correlations
is carried by quasi-particle pairs generated during the quench.
Experimental evidences for the light-cone effect were uncovered in cold atomic
gases~\cite{CheneauKuhr2012, LangenSchmiedmayer2013}, and in ionic
systems characterized by long range interactions~\cite{JurcevicRoos2014}. 
As first observed in \cite{CheneauKuhr2012}, a bosonic gas was initially prepared deep within the Mott insulator phase by the  
application of an optical lattice, then a sudden decrease of the intensity of the lattice potential brought
the system slightly closer to the superfluid phase, and the subsequent evolution of the correlations was
monitored and spatially resolved. Even for this non-integrable model, an approximate light-cone
like propagation was found as shown in Fig.~\ref{fig:Fig1} (dashed line). Deep within
the Mott insulator, the correlations were carried by the ballistic evolution of holons
and doublons, quasi-particles corresponding to holes and excess particles within an atomic Mott insulator.

While the evolution of isolated systems has been the topic of numerous investigations, very few
studies have attempted to clarify the influence of environmental couplings on the propagation of
correlations.
Works in this direction mainly focused on the behavior of correlations at long times showing,
within the Markovian limit, that correlations decay on a length scale set by the Lieb-Robinson velocity
and the system relaxation time~\cite{Poulin2010, NachtergaeleZagrebnov2011}, and that
an event horizon may emerge~\cite{Descamps2013, Descamps2015}.
Furthermore, in integrable models, the thermalization dynamics of correlations 
in the presence of white noise or starting from
finite temperature initial states were explored in \cite{MarinoSilva2012} and
\cite{BonnesLauchli2014} respectively. While the environmentally-assisted diffusive emergence of long-lasting
exotic pair correlations was investigated theoretically in \cite{BernierKollath2013}.
Despite these recent findings, very little is known about the propagation of experimentally
measurable correlations in dissipative strongly interacting systems.

To fill this gap, we focus here on the propagation of correlations in the one-dimensional
Bose-Hubbard model following an abrupt interaction change while in contact
with a memory-less environment which causes dephasing.
We study this system both numerically and analytically.
Numerically, we use the time-dependent matrix product state method
(t-MPS)~\cite{Vidal2003,DaleyVidal2004,WhiteFeiguin2004} simulating density operators via
purification~\cite{ZwolakVidal2004,VerstraeteCirac2004} building upon a very recent improvement where the number of
particles in the physical and auxiliary spaces are conserved separately. This technical development
is required to perform efficiently these numerically expensive simulations and reach the relevant time-scales.
Analytically, we describe the dynamics in terms of holon and doublon excitations. By contrast to
isolated systems where the equations of motion are exactly solvable within this rewritting in terms of
excitations, here due to the presence of dissipation further decouplings are needed
for the system to become analytically tractable. The technique developed is valid up to intermediate times
for sufficiently large interaction strengths. Within both approaches, we find that dissipation does not affect
all correlations equally. Single-particle correlations exhibit, up to intermediate times, a light-cone
like propagation as in isolated systems.
The presence of the dissipation effectively speeds up the propagation while reducing the coherence.
In contrast, for the equal-time two-point density correlations the dissipative heating becomes quickly dominant and
the initial ballistic propagation regime gives way to diffusion.
These drastic effects, which dissipation has on the propagation of
correlations, will strongly influence the formation and stability of various sought-after quantum states
which experimentalists are currently attempting to dynamically induce.

%
%
We study the dynamics of strongly interacting bosons
in a one-dimensional lattice under the effect of local dephasing noise. The evolution of the
density operator $\rhop$ is described by the master equation 
\begin{align}
\frac{d\rhop}{dt} = -\frac{\im}{\hbar} \spare{\Hop,\rhop} + \Dop (\rhop) \label{eq:master}.
\end{align} 
The first term on the right-hand side describes the unitary evolution due to the Bose-Hubbard Hamiltonian 
\begin{align}
  \Hop=-J\sum_j \pare{\adop_j\aop_{j+1} + \hc } +\frac{U}{2} \sum_j\nop_j\pare{\nop_j -1}, \nonumber
\end{align} 
while the second term describes the dephasing noise in Lindblad form
\begin{align}
  \Dop(\rhop) = \ga \sum_j \left(\nop_j \rhop \nop_j -\frac 1 2 \gpare{\nop_j^2,\rhop}\right), \nonumber
\end{align}
where $\adop_j$ ($\aop_j$) creates (annihilates) a boson at site $j$ while $\nop_j=\adop_j\aop_j$.
Here $J$ is the tunnelling constant, $U$ the interaction strength, $\ga$ the dissipation strength, and
$\gpare{\cdot,\cdot}$ is the anti-commutator between two operators. The Bose-Hubbard Hamiltonian presents,
at a critical ratio $\pare{U/J}_c$ and for integer lattice fillings, a transition between a coherent
superfluid phase, for $U/J<\pare{U/J}_c$, and an incompressible Mott insulator
phase for $U/J>\pare{U/J}_c$~\cite{BarmettlerKollath2014}. 
In one dimension and for constant filling $\nbar=\tr\pare{\rhop~\nop_j}=1$, the transition occurs at a
critical ratio $\pare{U/J}_c\approx 3.4$~\cite{KashurnikovSvistunov1996, KuhnerMonien1998}. 

The second term of Eq.~(\ref{eq:master}), $\Dop(\rhop)$, describes the effect of local dephasing, and,
in the cold atom context, can be the dominant contribution arising from spontaneous emission from the
optical lattice laser beams~\cite{PichlerZoller2010, GerbierCastin2010, PatilVengalattore2015, LueschenSchneider2017}.
This dissipator acts like a source of heat, generating excitations which increase local particle fluctuations and
eventually would drive, within this model, the system towards the infinite temperature state.
A stroboscopic application of the same quantum jump operator was
shown to change the transport of particles from ballistic to diffusive
as signalled by the evolution of local density~\cite{KesslerMarquardt2012}.

We consider here a system at unit filling, $\nbar=1$, initially prepared in
the groundstate of the Bose-Hubbard model at $U/J \rightarrow \infty$,
the atomic Mott insulator state.  We investigate the propagation of
correlations after an abrupt change of the interaction strength to a
large but finite $U/J$ value such that the corresponding groundstate would be Mott-insulating.
As the interaction quench takes place, the system is coupled to an environment and
starts to undergo dephasing.

%
%
To gain a deeper analytical understanding of the initial dynamics, we reformulate the
description of this system in terms of fermionic quasi-particles. For large final
interaction strengths, this method was shown to provide accurate quantitative
predictions and qualitative insights into the non-equilibrium dynamics of closed systems
following a sudden quench~\cite{CheneauKuhr2012, BarmettlerKollath2012}.
This approach is based on the realization that, for the isolated system and in the
large interaction limit, the local occupation of a site ranges almost
exclusively between $\nbar-1$ and $\nbar+1$.
Hence, in one-dimension, making use of the Jordan-Wigner transformation, two types of auxiliary
fermions can be introduced, $\cop_{j,\pm}$, such that 
\begin{align}
  \adop_j=\sqrt{\nbar+1}~Z_{j,+}~\cdop_{j,+} + \sqrt{\nbar}~Z_{j,-}~\cop_{j,-} \nonumber
\end{align}      
where $Z_{j,+}=e^{\im\pi\sum_{\sigma, l<j}\nop_{l,\sigma}}$, $Z_{j,-}=Z_{j,+}e^{\im\pi\nop_{j,+}}$, $Z^\dagger_{j,\pm} = Z_{j,\pm}$ and
the ``$+$'' and ``$-$'' fermions are doublons and holons respectively.
The rewriting of excitations as fermions ensures that multiple doublons (or multiple holons) do not occupy the same site.
The Bose-Hubbard Hamiltonian then reduces to
\begin{align}
  \Hop_\text{f} =
  \sum_j \Pop & \left[-J\pare{\nbar +1}\cdop_{j,+}\cop_{j+1,+} - J\nbar~\cdop_{j,-}\cop_{j+1,-} \right. \nonumber \\
  &~ -J\sqrt{\nbar\pare{\nbar +1}} \pare{\cdop_{j,+}\cdop_{j+1,-} - \cop_{j,-}\cop_{j+1,+}}  \nonumber \\ 
    &~ \left. +\frac{U}{4} \pare{\nop_{j,+} +\nop_{j,-}} + \hc\right]  \Pop, \nonumber
\end{align} 
where the projector $\Pop = \prod_j (1-\nop_{j,+}\nop_{j,-})$ ensures that a doublon and a holon do not
occupy the same site. In the following we set $\Pop=1$
rendering the Hamiltonian $\Hop_\text{f}$ quadratic and readily diagonalizable~\cite{CommentonValidity}.
However, the dissipator written in terms of fermionic quasi-particles, using $\nop_j=\nbar+\nop_{j,+} - \nop_{j,-}$ and
neglecting terms in $n_{j,+} n_{j,-}$ which are null in the presence of $\mathcal{P}$, is still not quadratic
\begin{align}
  \Dop_\text{f}(\rhop) =~ & \ga \sum_{j,\sigma=\pm} \left( \nop_{j,\sigma}\rhop\nop_{j,\sigma}
  -\frac{1}{2} \{\nop_{j,\sigma},\rhop\}  - \nop_{j,\sigma} \rhop \nop_{j,\bar{\sigma}}\right), \nonumber
\end{align}   
with $\bar{\sigma}=-\sigma$. Nevertheless, a closed set of differential equations can be
obtained for the correlations
$\Cpm(d) = 1/L \sum_{j} \la \cop_{j+d,+} \cop_{j,-} \ra$
and
$K(d) = 1/L \sum_{j} \la \cdop_{j+d,+} \cop_{j,+} \ra = 1/L \sum_{j} \la \cdop_{j+d,-} \cop_{j,-} \ra$.
Due to the choice of the initial state, and as $K^*(d)=K(-d)=K(d)$ and $\Cpm(d)=-\Cpm(-d)$, we get       
\begin{align}
\hbar\frac{d}{dt} \Cpm(d) = &  -\hbar\ga~\Cpm(d) -\im U~\Cpm(d)  \nonumber \\    
 &  -\im J \sqrt{\nbar(\nbar+1)} \pare{\delta_{d,-1} - \delta_{d,1}} \nonumber \\    
 &  -\im 2J \sqrt{\nbar(\nbar+1)} \spare{K(d-1) - K(d+1)} \nonumber \\  
 &  +\im J (1+2\nbar) \spare{\Cpm(d-1) + \Cpm(d+1)}; \nonumber \\
\hbar\frac{d}{dt} K(d) = &  -\hbar\ga (1-\delta_{d,0}) K(d) \nonumber \\    
 &  +\im J \sqrt{\nbar(\nbar+1)} \left[ \Cpm(d-1) - \Cpm(d+1) \right. \nonumber \\      
 &\left. +~\Cpm^*(-d+1) - \Cpm^*(-d-1) \right]. \nonumber          
\end{align}
At $O(J/U)$, we obtain an explicit solution for $\Cpm(d,t)$:
\begin{align}
\Cpm(d,t) = & \frac{J}{U} \left[\sqrt{\nbar(\nbar+1)} \; \frac{1+\im(\hbar\ga/U)}{1+\left(\hbar\ga/U\right)^2}
\pare{\delta_{d,-1} - \delta_{d,1}} \right. \nonumber \\ 
+ & \left. \im (-1)^d \frac{\sqrt{\nbar(\nbar+1)}}{1+\left(\hbar\ga/U\right)^2} \; e^{-\ga t} f(d,t) \;g(d,t)   \right]  \label{eq:cpmsol}        
\end{align}
where $f(d,t)=J_{d+1}(\tilde{J}t) + J_{d-1}(\tilde{J}t)$, with $\tilde{J} = 2(1+2\nbar)J/\hbar$,
$J_n$ being the $n^\text{th}$ order Bessel function of the first kind,
and $g(d,t)= \pare{\hbar\ga/U-\im}\sin\pare{Ut/\hbar + \pi d /2} +  \pare{\im\hbar\ga/U+1}\cos\pare{Ut/\hbar + \pi d /2}$~\cite{Commentd1}.
In comparison, the leading term in the solution for $K(d,t)$ is of order $(J/U)^2$.
%
%

%
%
As shown in Fig.~\ref{fig:Fig1}, for a dissipative coupling strength $\hbar\gamma = 0.4J$,
the behavior of $\Cpm$ is non-trivially affected by the presence of an environment.
The (blue) points are results from t-MPS simulations \cite{MPS_details} while the (grey) full line, closely following these points, is the
solution obtained within the quasi-particle picture.
In the presence of dissipation, these correlations still propagate
ballistically up to intermediate times while their coherence is reduced by $e^{-\ga t}$
as illustrated by the envelope of the analytical solution, the (green) full line.
However, as evidenced by the analytical expression, the effect of dissipation cannot be reduced
to a simple exponential damping since the dissipation strength $\gamma$ also enters through
$g(d,t)$ as a prefactor to the oscillation term. In addition, the (red) solid line tracks the
position of the first maximum of the envelope of $\Cpm$ for $\hbar\gamma = 0.4J$. 
For a given distance $d$, the first maximum appears at
an earlier time compared to the $\gamma=0$ case signalling an effective speed-up of the
propagation of single-particle correlations. To support this statement, the time at which
the first maximum reaches a given distance is reported for various dissipation strengths
in the inset of Fig.~\ref{fig:Fig1}. The observed increased velocity is a result of the
interplay of the linear propagation of the envelope of $\Cpm$ and its exponential decay. 

\begin{figure}
\includegraphics[width=0.9\columnwidth]{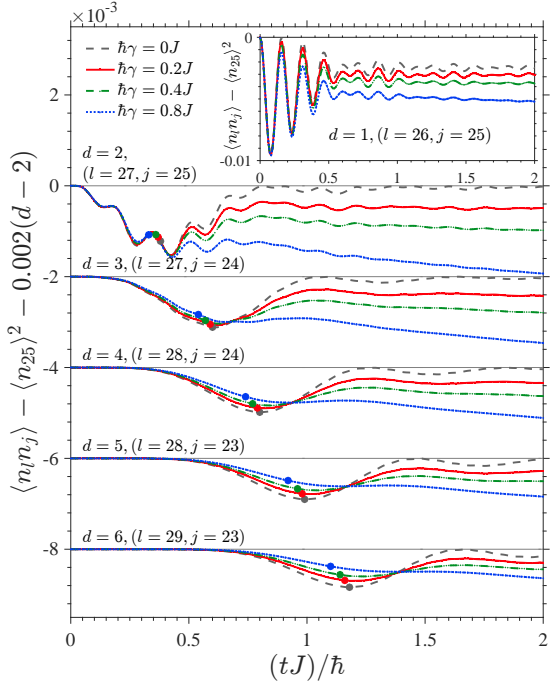}
\caption{(color online) Evolution of two-point density correlations following a sudden
  interaction change from an atomic  Mott insulator to $U=40J$ for various
  distances and dissipative strengths (the correlations are shifted by $-0.002(d-2)$).
  The large dots mark the times at which the first maximum
  of $\Cpm$ reaches a given distance. \label{fig:Fig2}}
\end{figure}

As discussed earlier, the dephasing does not affect all correlations equally. While,
under both unitary and dissipative dynamics, the propagation
of $\Cpm$ is approximately ballistic, the evolution of the two-point density
correlator is strongly altered by the coupling to an environment.
Dissipation not only reduces coherence, but also heats up the system and increases number
fluctuations leading to the proliferation of doublon-holon pairs and
to the creation of even higher energy excitations corresponding to larger site occupancy. These effects
are evidenced by the evolution of the density correlations plotted versus
time for various dissipation strengths shown in Fig.~\ref{fig:Fig2}.
At short times, the propagation follows approximately the light-cone like regime, known from
isolated systems, with an additional damping $e^{-2\ga t}$.
This is in agreement with results obtained from the quasi-particle picture. Within
a mean-field decoupling of the density correlations, we obtain, to leading order in $J/U$,
$\frac{1}{L}~\sum_j~\la \nop_{j+d} \nop_{j} \ra_\text{MF} \approx -2 |\Cpm(d)|^2$ for $d>1$. 
However, for larger times, heating
dominates the correlations and obliterates the light-cone structure. On Fig.~\ref{fig:Fig2} we indicate,
using large dots, the times at which the first maximum of $\Cpm$ reaches a given distance. For $d>2$,
one sees that for $\ga = 0$, the extrema of the single-particle and density correlations reach
a given distance at the same time. For finite dissipation strengths the two extrema start
to deviate as the maximum of the single-particle correlations effectively speeds up while the minimum of
the density correlations shifts towards larger times. For sufficiently large $\ga$
or distances, this minimum is overwhelmed by the effect of heating rendering impossible its identification.
Therefore, we find that at intermediate times the original ballistic propagation is replaced
by a different propagation regime.

\begin{figure}
\includegraphics[width=0.9\columnwidth]{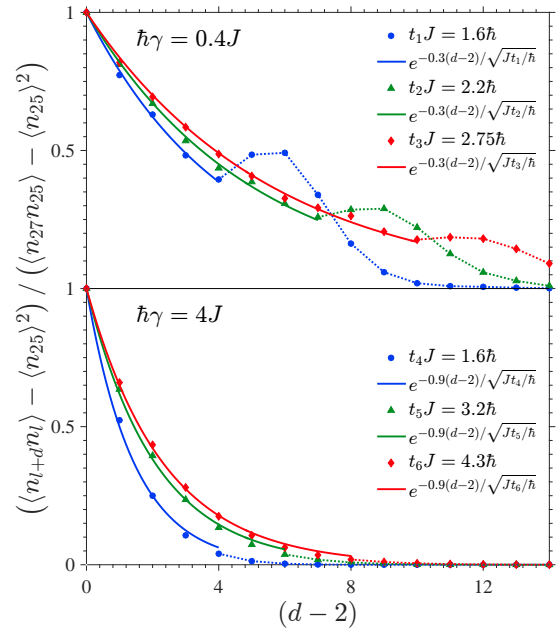}
\caption{(color online) Normalized density correlations as a function of distance for various times
  and two dissipative strengths. The full lines are fits to the
  function $\exp(-\alpha(d-2)\sqrt{\hbar/(Jt)})$, $\alpha$ is the only free parameter. The dashed lines
  are solely guides to the eye. \label{fig:Fig3}}
\end{figure}

%
%
In order to understand this emerging regime, we plot
in Fig.~\ref{fig:Fig3} these correlations as a function of distance for various times and two different
dissipation strengths. For weak dissipation, $\hbar\ga = 0.4J$, where the light-cone peak is still
distinguishable, we find that after the passage of this maximum the correlations propagate, for $d>2$, as
$\exp(-\alpha(d-2)\sqrt{\hbar/(Jt)})$, where $\alpha$ is a free fitting parameter, hinting
that the initial ballistic propagation has given
way, at intermediate times, to diffusion. 
The emergence of this different propagation
regime is also observed for larger $\ga$ as shown on the lower panel of Fig.~\ref{fig:Fig3}.
In this case, the light-cone peak is totally suppressed but one still sees that the size of the
region over which correlations propagate diffusively increases as a function of time. The values of
the fitting constant $\alpha$ agree for different times even though, at each time, it is
extracted independently. Thus, $\alpha$ appears to be dependent
mainly on the strength of the dissipative coupling.

To summarize, we analyzed the propagation of correlations in a strongly interacting Bose gas
confined to an optical lattice after a sudden interaction change while the system is in contact with an environment.
We found that the presence of dissipation does not affect the density and single-particle
correlations in the same manner. Compared to the case of an isolated system, dissipation effectively
speeds up the light-cone like propagation of single-particle correlations although their signal
is suppressed over time. In contrast, for two-point density correlations, the initial ballistic propagation
is rapidly overwhelmed by heating and enters a diffusive regime. The system studied here is experimentally
realizable using the same state of the art experimental techniques developed in \cite{CheneauKuhr2012}
to investigate an isolated quench, while only the dissipative coupling would need to
be strengthened~\cite{PatilVengalattore2015, LueschenSchneider2017}.
This coupling could also be enhanced by the application of an additional laser fluctuating both
in space and time.
We expect the effects uncovered in this study to be observable over a wide regime of
parameters, to arise in different systems with Markovian noise, and to have important implications on the dynamical
generation and evolution of complex states.
Future studies might address how the
propagation of correlations is affected by couplings to non-Markovian environments with a particular focus
on solid state systems where, for example, phonons can constitute an important dissipative channel~\cite{GiannettiMihailovic2016}.

{\it Acknowledgments:} We thank S. Diehl, A. L\"auchli and J. Marino for fruitful discussions.
We acknowledge support from Singapore Ministry of Education, Singapore Academic Research Fund Tier-I
(project SUTDT12015005) (D.P.) and DFG (TR 185 project B3 and Einzelantrag) (C.K.).

\end{document}